\documentstyle[editedvolume,epsf]{crckapb}

\newcommand{\czw}{$^{\rm 12}{\rm C}$}

\def\etal{{et\thinspace al.}\ }

\begin{opening}
\title{Internal mixing and surface abundance of $\;\;\;$ [WC]-CSPN}
\author{F. Herwig}
\institute{Universit\"at Potsdam, Germany \\
        fherwig@astro.physik.uni-potsdam.de}
\end{opening}
\runningtitle{Mixing and surface abundance}
\begin{document}

\begin{abstract}
Recent advances in constructing stellar evolution models of
hydrogen-deficient post-asymptotic giant branch (AGB) stars are presented.
Hydrogen-deficiency
can originate from   mixing and subsequent convective burning of
protons in the deeper layers during a thermal pulse on the post-AGB (VLTP).
Dredge-up alone may also be
responsible for hydrogen-deficiency of post-AGB 
stars. Models of the last thermal pulse on the AGB with very small
envelope masses have shown efficient third dredge-up. The
hydrogen content of the envelope  is diluted sufficiently  to produce
H-deficient post-AGB stars (AFTP).
Moreover, dredge-up alone may also cause H-deficiency during the
Born-again phase (LTP).
During the second AGB phase a
convective envelope develops. A previously unknown
lithium enrichment at the surface of Born-again stellar models may be
used to distinguish between objects 
with different post-AGB evolution.
The observed abundance ratios of C, O and He
 can be reproduced by all scenarios if an AGB starting model
with inclusion of 
overshoot  is used for the
post-AGB model sequence. 

An appendix is devoted to the numerical methods for models
of proton capture nucleosynthesis in the He-flash convection zone during a 
thermal pulse. 
\end{abstract}

\section{Introduction}
\begin{figure}[tp]
\epsfxsize=\textwidth 
 \epsfbox{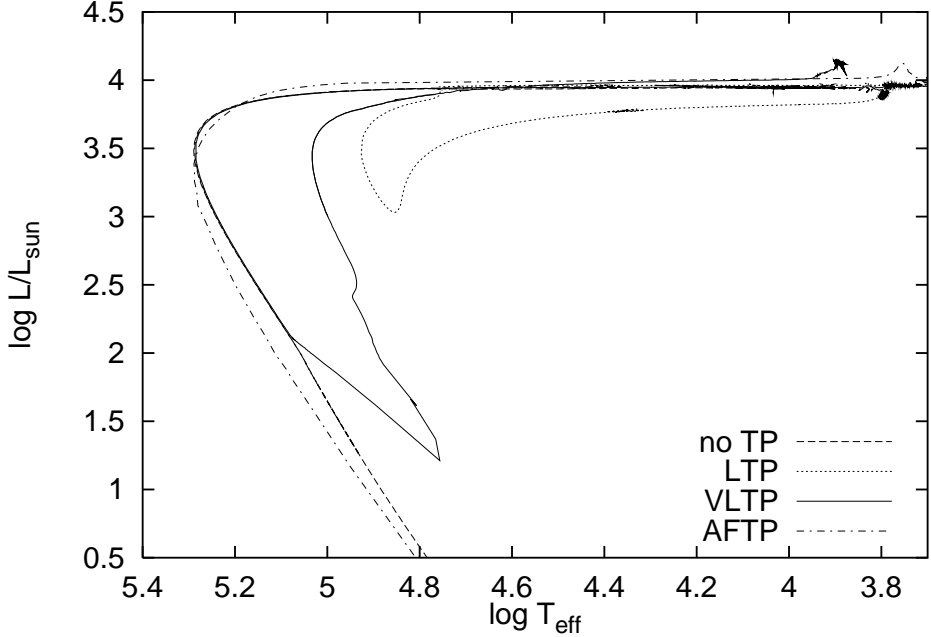}
\caption{ \label{fig:POST-AGB-HRD-ALL} 
Evolutionary post-AGB tracks in the HRD for the four 
scenarios discussed in the text. The stellar mass is always 0.6M$_\odot$.
}
\end{figure} 
\begin{figure}[tp]
\epsfxsize=\textwidth 
 \epsfbox{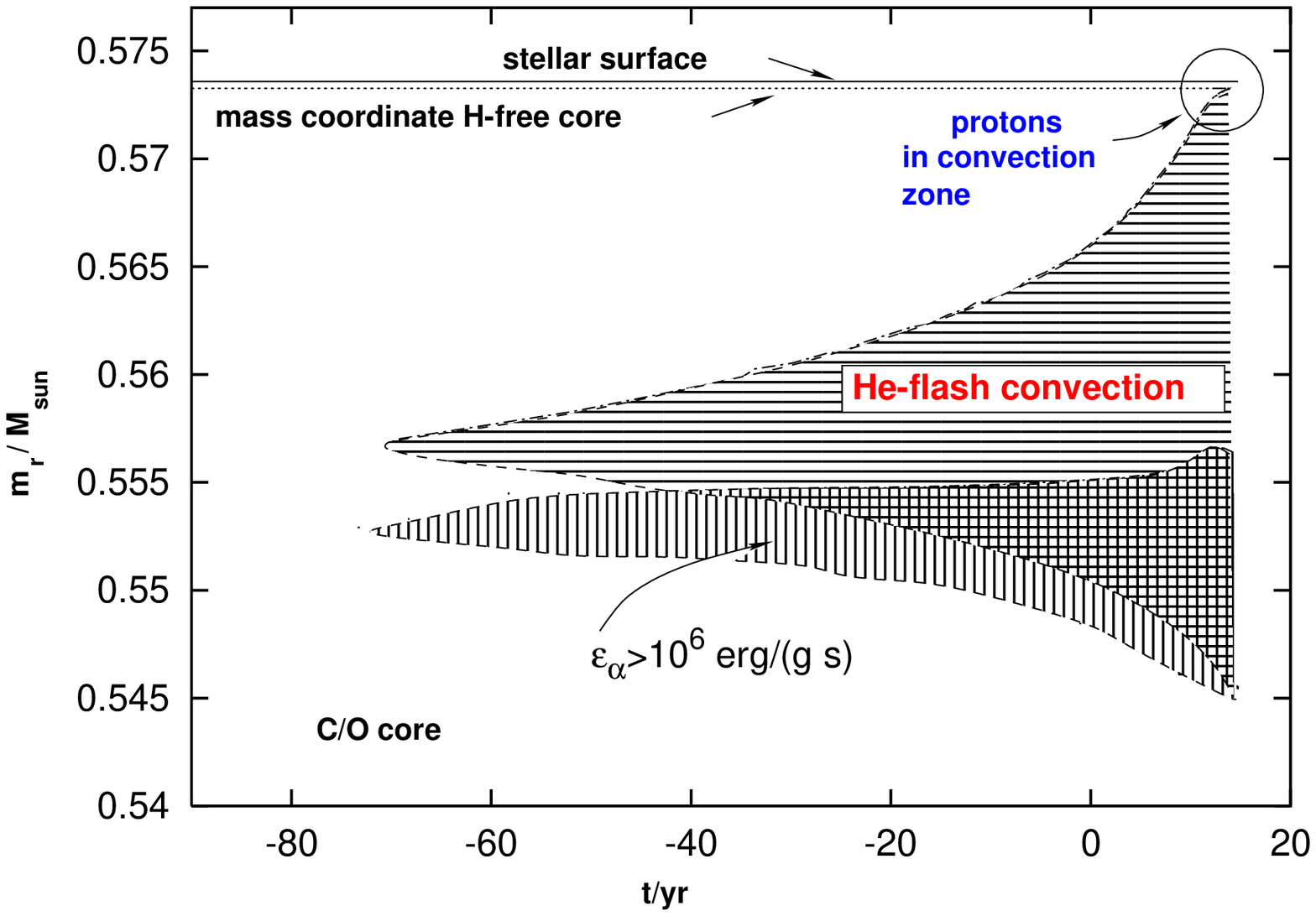}
\caption{ \label{fig:KIPPI} 
Internal evolution of the  He-flash convection zone towards a post-AGB 
TP. Horizontal hatch is used for the growing He-flash convection
zone. It  is driven by the increasing energy production due to
He-burning. The latter  is marked by vertical hatches.
}
\end{figure}
The problems of interpreting the evolutionary origin of H-deficient
post-AGB stars in general and the [WC]-central stars of planetary
nebulae (CSPN) in particular are 
 described in this volume, e.g.\ by Koesterke and
by Werner. These
peculiar stars are found at positions in the HRD which correspond to
post-AGB evolutionary sequences of low- and intermediate 
\enlargethispage*{4.5cm}

\vspace{3cm}
\noindent
\begin{minipage}[h]{\textwidth}
\hrule
\medskip
\texttt{Workshop on ``Low mass Wolf-Rayet stars: origin and
evolution'', 1999, Amsterdam, to appear in ApSS, Kluwer.}
\end{minipage}
\pagebreak

mass
stars (Fig.\,\ref{fig:POST-AGB-HRD-ALL}).
Hydrogen is underabundant at the surface
and a closer look at the various
observational analysis reveals a diverse picture.
H-rich stars with practically solar H-abundance,
so-called hybrid objects with substantial fractions of typically more 
than $10\%$ of H, H-poor with less than a few percent H and
definitely H-free descendants of AGB stars like the DO-white dwarfs (WDs)
have been found. With new stellar evolution models
we have now
identified several possibilities for the origin of H-deficiency of
post-AGB stars which make different predictions for the degree of
H-depletion and other elements.
The recent progress  in modeling these objects are either related to
the mixing and simultaneous burning of the remaining envelope
hydrogen or to the  dredge-up (DUP). The new results are based on 
progenitor AGB models with overhooting.

The first possibility is
the so-called \emph{very late thermal pulse} (VLTP) and has been first 
suggested by Fujimoto (1977).
Secondly, H-deficiency may
occur after the last thermal pulse on the AGB.
When the remaining envelope mass is coincidently very small at that
last TP, one final dredge-up phase can significantly reduce the
H-abundance and we have phrased the term \emph{asymptotic
giant branch final thermal pulse} (AFTP) for this situation. In the
third case,
DUP occurs during the Born-again
evolution which follows a post-AGB thermal pulse (either late or very
late). If the model has not become H-deficient already due to combined 
mixing and burning during a very late thermal pulse, then
 it will become H-deficient due to DUP during its
second AGB evolution after a \emph{late thermal pulse} (LTP, see Bl\"ocker, this  
volume).

The models presented in this paper have been computed with our stellar 
evolution code as described by Bl\"ocker (1995) and extended with
respect to mixing (Herwig \etal1997) and coupled nuclear burning
\mbox{(Herwig \etal1999)}. For the sake of numerical ease the 
model sequences have been computed with older
opacities (Cox \& Stewart, 1970) and a mixing length parameter of
$\alpha_{MLT}=3$. While qualitatively the results presented in this
paper should not be affected by this choice there are possibly details 
which are depending on these assumptions.

\section{Decision making: no-TP, AFTP, LTP or VLTP?}
\label{kap:decision}
The evolution of TP-AGB stars is ruled by an internal clock and the TPs 
which occur repeatedly after almost equidistant time intervals set
the pace at which this TP clock runs.
Whether or not and to which degree an AGB star will show H-deficiency later on 
depends on the phase of the TP pulse cycle  at which the star leaves
the AGB. It defines
the TP time until the next TP would occur during an 
undisturbed TP cycle. At this moment another clock starts to run at
the pace at 
which the remaining envelope is reduced by H-burning and mass
loss. Moreover, when the star leaves the AGB there is  the
He-burning clock starting as well, which measures the time until the
object has become 
a WD and all  nucleosynthesis has practically ceased. This
time is always longer than the H-burning time.

The post-AGB fate of a star depends on the ratio of these three
time scales. If the star leaves the AGB at an early pulse cycle phase
then the TP time is much longer than the He-burning time and when the
next thermal would be due the star has already become a white
dwarf. This  \emph{no-TP} case is the most common.
At a larger phase the TP time is shorter than the
He-burning time but still larger than the H-burning time. In that case 
a thermal pulse (VLTP) occurs when the evolutionary path  in
the HRD is 
already directed towards the WD cooling track.
In the VLTP case the He-flash convection zone is engulfing
the envelope and the hydrogen in the outermost layers is mixed
downwards and burned (Fig.\ \ref{fig:KIPPI}, circle).
If the star leaves the AGB at an even
larger phase the TP time is smaller than the H-burning
time. A thermal pulse (LTP) will occur and the situation will again be 
similar as displayed in Fig.\ \ref{fig:KIPPI}. But this time the
hydrogen burning shell has not yet ceased and no mixing of envelope
material into the deeper layers will be possible (Iben, 1976). The surface
abundance is undisturbed for another while until DUP during the
Born-again evolution leads to efficient H-depletion. Finally, one more
possibility leads to a chemically peculiar evolution. If the star
leaves the AGB immediately after a TP (AFTP) the remaining envelope
mass must be already very small. DUP during this phase then
leads to H-deficiency already during the first (and in this case only) 
post-AGB evolution.

Thus, the post-AGB fate is determined by the interpulse phase at which 
the post-AGB evolution starts. This happens when mass loss has
reduced the envelope mass below a core-mass dependent critical
value. Therefore, the segmentation of the post-AGB evolution
is a statistical process impressed by the mass loss as a function of phase.
\begin{figure}[tbp]
\epsfxsize=\textwidth 
 \epsfbox{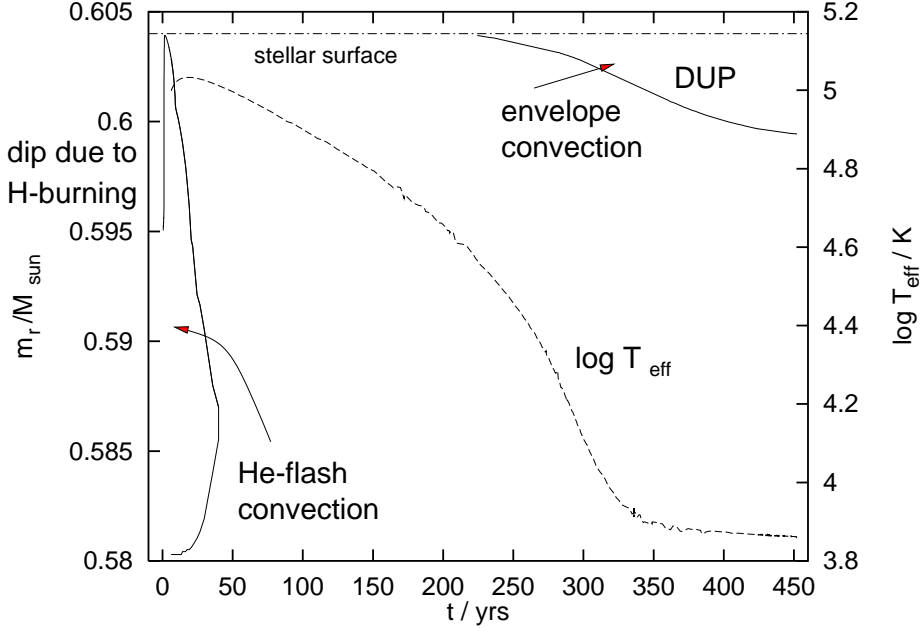}
\caption{ \label{fig:DUP-6.III.fig} 
Internal Born-again evolution after a VLTP. The solid lines give the
position of the boundaries of convection zone and the dash-dotted line 
indicates the position of the stellar surface (left scale).
The dashed line gives the effective temperature (right scale) and
shows how the model cools back into the AGB regime after the VLTP.
The zero point is set to the peak He-luminosity
of the pulse. The dip at the top of the He-flash  convection zone at
$t=0 {\rm yr}$ is
caused by the sudden and rapid energy release from H-burning and is
identical with the prominent one year long dip displayed in Fig.\,2
in Herwig \etal(1999).
 }
\end{figure} 
\begin{figure}[ptb]
  \begin{center}
  \epsfxsize=0.8\textwidth
   \hspace*{0.8cm}
   \epsfbox{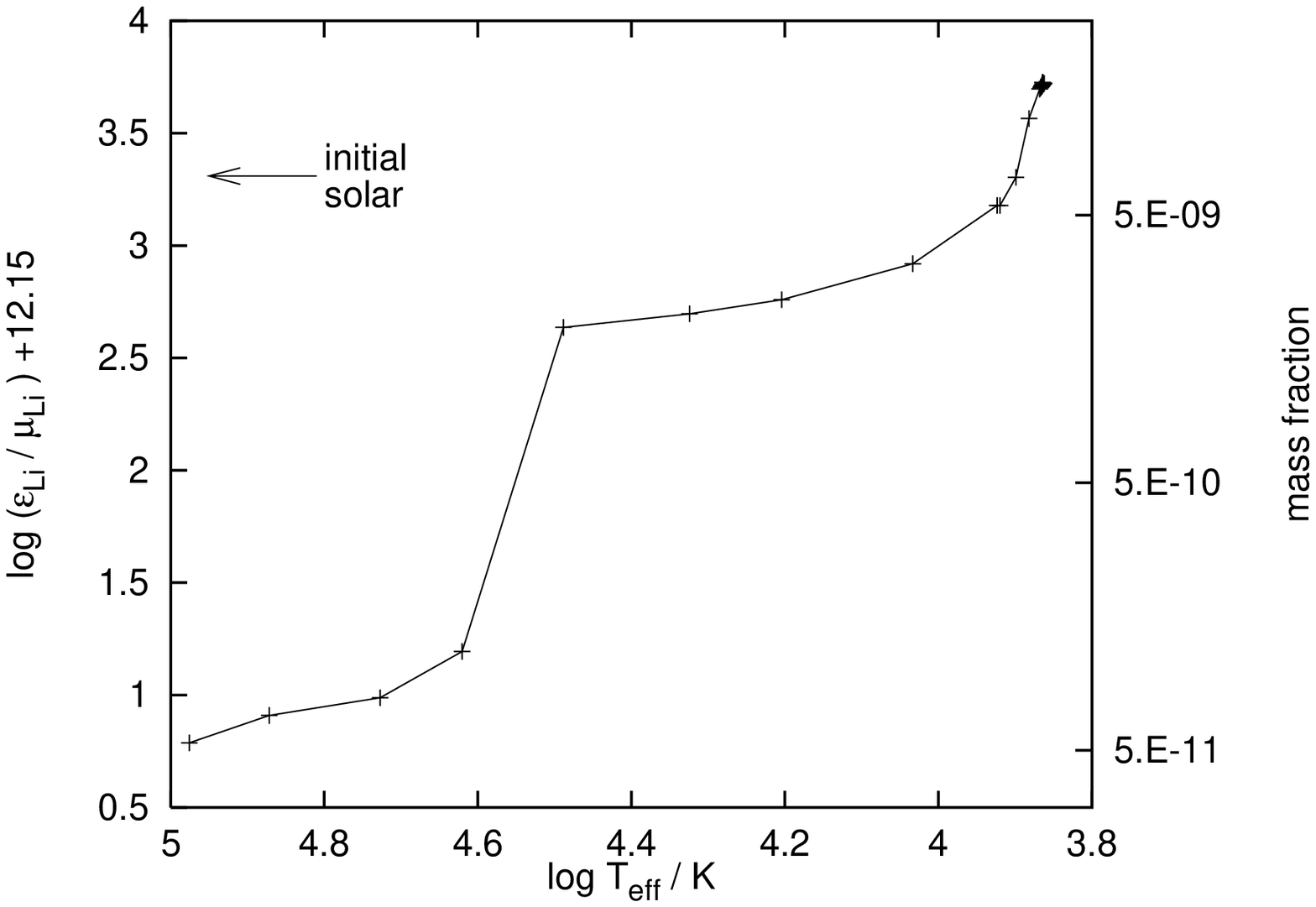}
  \caption{ \label{fig:Li-6.III} 
     Evolution of lithium abundance in the convective zone below the surface during the evolution after
    a VLTP back to the AGB domain in the HRD (see
    Fig.\,\ref{fig:DUP-6.III.fig}).  }
  \end{center}
\end{figure} 

\section{Post-AGB stars suffering a very late thermal pulse}
\label{kap:VLTP}
The VLTP scenario has been investigated by Iben \& MacDonald (1995)
with a semi-analytic treatment of burning protons as they are entering 
the He-flash convection zone (Fig.\,\ref{fig:KIPPI}).
Recently Herwig \etal(1999) have
improved on this study by employing a fully coupled solution  of
a nucleosynthesis network and a diffusion equation for each
isotope. The numerical method is described in the appendix of this
paper.
Hydrogen is entirely destructed but no $^{14}{\rm N}$ is produced. Instead
protons ($\sim {5}\cdot 10^{-5}{\rm M}_\odot$) are efficiently
transformed first into $^{13}{\rm C}$ which serves as a neutron source 
via the reaction chain $^{12}{\rm C}({\rm
p},\gamma)^{13}{\rm N}({\rm e}^+ \nu)^{13}{\rm C}(\alpha,{\rm
n})^{16}{\rm O}$. Also we assumed that overshooting was operating in
the progenitor AGB star and thereby the large oxygen mass fraction
(our model: He/C/O = 0.38/0.36/0.22) 
observed in many H-deficient post-AGB stars has been reproduced.

Following the evolution back to the second AGB phase has revealed that 
Born-again AGB stars which descend from AGB models with overshoot
develop a convective envelope while the outer layers are expanding
and cooling. In our VLTP descendant the envelope convection develops
when $\log T_{\rm eff} < 4.6$ 
(Fig.\,\ref{fig:DUP-6.III.fig}). As the star cools further
envelope convection zone deepens with
respect to mass, leading to  {\bf dredge-up}. We choose the term
\emph{very late dredge-up} (VLDUP) for this event because this process 
after a VLTP has distinctly different properties compared to the well
known third DUP. The VLDUP
has not been reported from previous model sequences evolved from
AGB progenitors  without
overshoot. In the new models the efficient dredge-up  is related
to the
larger TP peak He-burning luminosity due to overshoot from below
the He-flash convection zone. Additionally the decreased intershell helium
abundance compared to models without overshoot favours a deep envelope 
convection.
 The neutron-capture nucleosynthesis (which has not
yet been followed in detail)  products which will form in the
He-flash convection zone after the ingestion of protons will now show
up at the  
surface. Another effect of this VLDUP is the significant
enrichment of {\bf lithium} (Fig.\,\ref{fig:Li-6.III}).
After a VLTP lithium is present in the intershell and mixed to the
surface by dredge-up. It is the result of $\beta$-decay of $^7{\rm
Be}$ which has formed in 
the intershell by $\alpha$-captures of $^3{\rm He}$.  $^3{\rm He}$
enters into the He-flash convection zone during the VLTP together with hydrogen 
from the envelope.
\begin{figure}[ptb]
\epsfxsize=\textwidth 
 \epsfbox{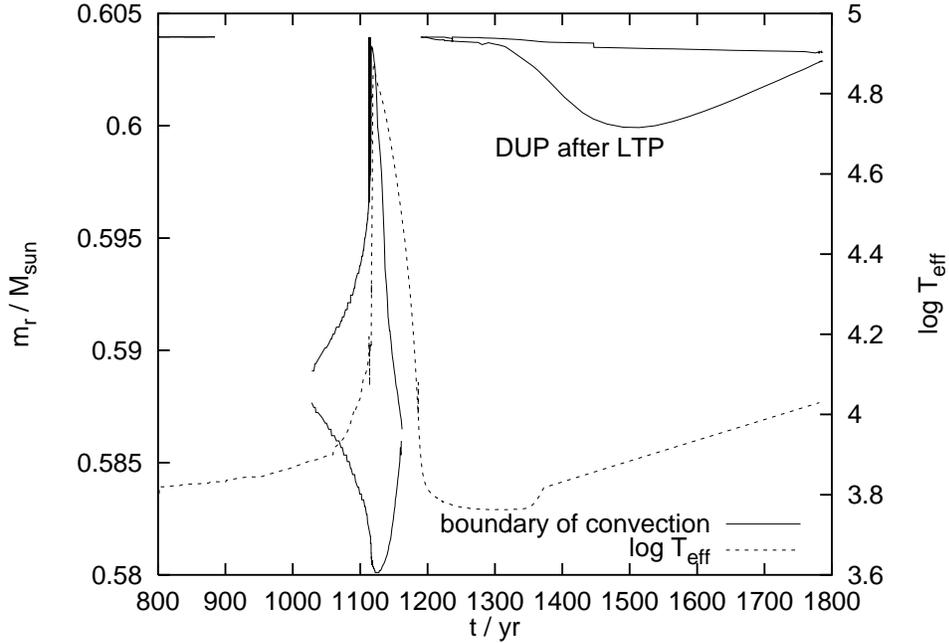}
\caption{ \label{fig:KIPP-TEFF-LTP.ps} 
Internal evolution of Born-again  post-AGB model sequence after a LTP.}
\end{figure}
\begin{figure}[htp]
\epsfxsize=\textwidth 
 \epsfbox{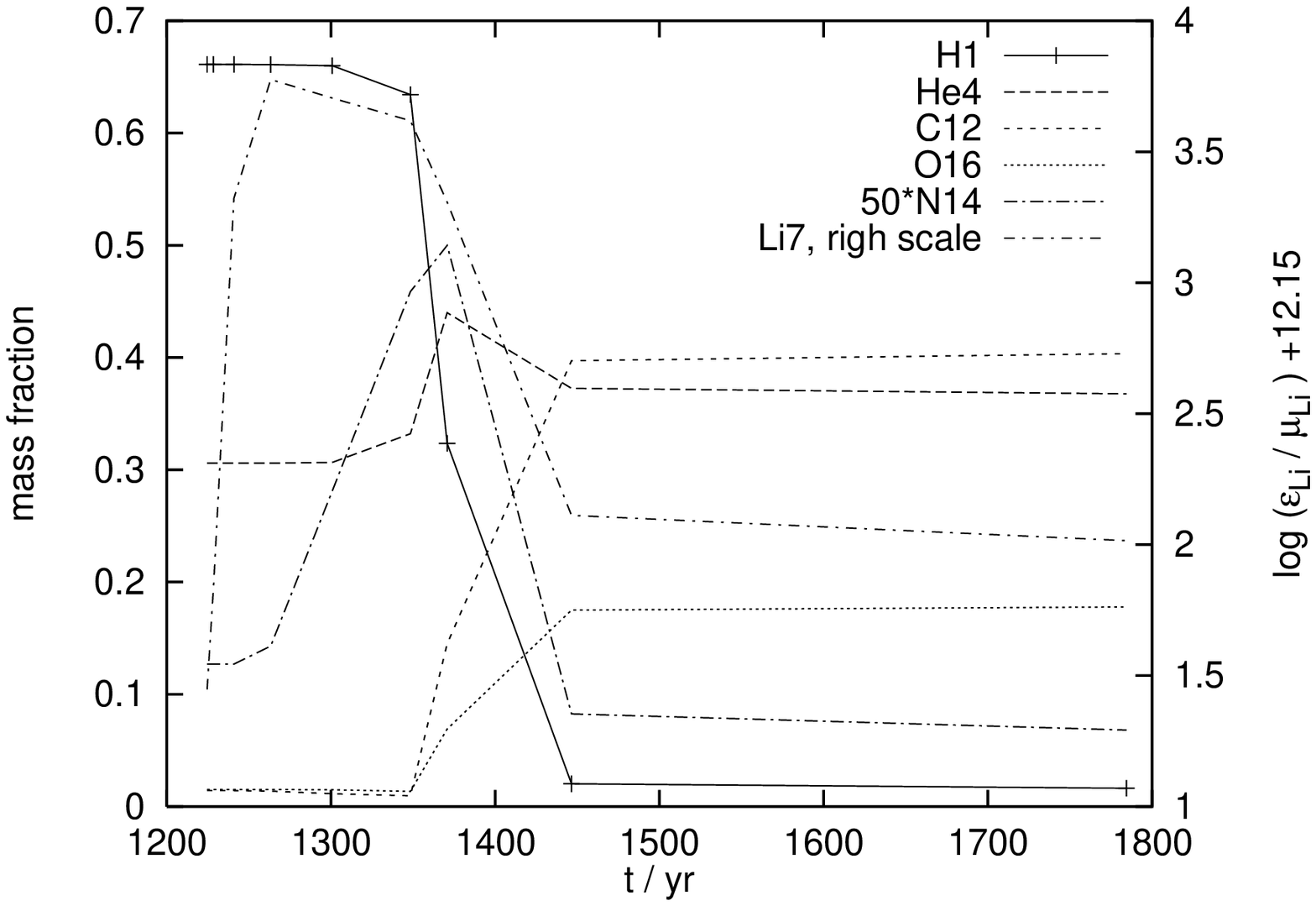}
\caption{ \label{fig:ABUND} 
Surface abundance during Born-again evolution after LTP. Lithium
abundance right scale.}
\end{figure}

\section{Post-AGB stars suffering a late thermal pulse}
Contrary to the case of the VLTP the surface abundance remains
undisturbed by the LTP itself. However, H-deficiency is caused by
dredge-up during the following Born-again phase.
The outer layers react to the enormous
energy release by He-burning during the LTP by a very fast evolution
back to the AGB. At $\log T_{\rm eff} \sim 4.0$ the outer layers
become convectively unstable,  similar to the  
previously described evolution 
after the VLTP (Sec.\,\ref{kap:VLTP}). The internal evolution is
displayed in Fig.\,\ref{fig:KIPP-TEFF-LTP.ps}. During the coolest stage of
the Born-again evolution again dredge-up is found. This
\emph{late DUP} (LDUP) does  cause a substantial abundance
change because the  remaining envelope mass is only $<10^{-4}{\rm
M}_\odot$. The surface abundance evolution changes as the bottom of
the envelope convection proceeds into the interior where it passes the 
various sites of nuclear burning (Fig.\,\ref{fig:ABUND})\footnote{Close comparison of
Fig.\,\ref{fig:KIPP-TEFF-LTP.ps} and \ref{fig:ABUND} show
an apparently unsynchronized behaviour of DUP and abundance change. It 
is related to a split of the envelope convection which probably is caused
by the opacities.} 
.
Immediately after the convective zone has formed it  enters into a
lithium-rich region. It originates from the $^7$Be pocket which forms
due to the partial pp-chain operation at the cool side of the H-shell.
Then the  convective region proceeds through the fully H-burned region 
and a corresponding peak in the helium and nitrogen abundance can be
seen in Fig.\,\ref{fig:ABUND}. Finally the dredge-up proceeds into the
former intershell region.
Now the intershell abundance shows up at the surface as it has
been formed cumulatively over the entire previous TP-AGB evolution
with overshoot. At this stage which is reached shortly after the
lowest temperature during the Born-again evolution the surface
abundances of this LTP model are $X_{\rm H} = 0.02$, $X_{\rm He} =
0.37$, $X_{\rm C} = 0.40$, $X_{\rm O} = 0.18$.

\begin{figure}[htp]
\epsfxsize=\textwidth 
 \epsfbox{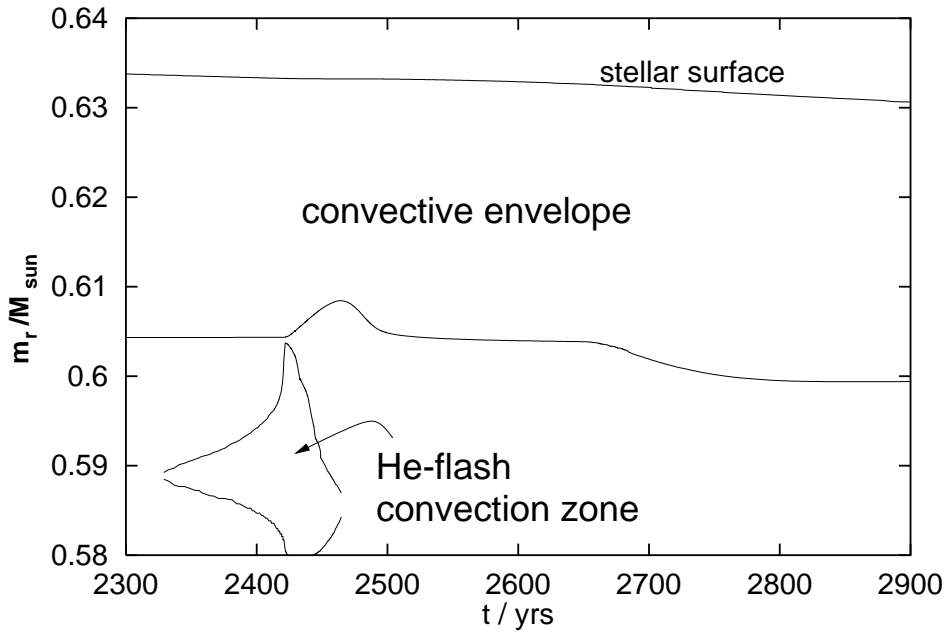}
\caption{ \label{fig:KIPP-5.II.A.fig} 
An AFTP model sequence: Evolution of mass coordinate
of convective boundaries and stellar surface during the last TP on the 
AGB (see text, case 1). DUP starts at $t \sim 2670 yr$.}
\end{figure} 

\section{The AGB final thermal pulse}
The third evolutionary channel also causes  H-deficiency by dredge-up
but this time the dredge-up occurs after the last (or final) thermal
pulse  on the AGB. This AFTP is characterized by a very low 
remaining envelope mass and it has, due to overshoot, a larger energy
production by He-burning compared to models without overshoot.
We have computed two AFTP cases which have been started from the same
AGB model sequence as the other post-AGB tracks. The mass loss has
been tuned in such a way that the envelope mass after the TP and
before the succeeding DUP was $3\cdot10^{-2}{\rm M}_\odot$ and
$4\cdot10^{-3}{\rm M}_\odot$ respectively (see
Fig.\,\ref{fig:KIPP-5.II.A.fig} for case 1). In both cases the
dredged-up mass was about $6\cdot10^{-3}{\rm M}_\odot$. 
At the end of this AGB final DUP (AFDUP) the surface abundance was
$X_{\rm H} = 0.55$, $X_{\rm He} =
0.31$, $X_{\rm C} = 0.07$, $X_{\rm O} = 0.04$ in case 1 and
 $X_{\rm H} = 0.17$, $X_{\rm He} =
0.33$, $X_{\rm C} = 0.32$, $X_{\rm O} = 0.15$ in case 2. 

\section{Conclusions}
Table \ref{tab:class} gives an overview of the H-abundances according
to our current understanding
of the formation of H-deficient post-AGB stars. The
H-abundances given are most uncertain for the AFTP case because of
poorly known input physics like the variation of the mass loss during
the thermal pulse. If the mass loss during the final AGB interpulse
period is constant and if only small modulations of the mass loss rate
during and immediately after the AFTP are assumed then the rate of
AFTP descendents should be only a few percent and presumably smaller
than the fraction of LTP descendants which can be estimated to be
somewhere between $5\%$ and $10\%$. An estimate for the VLTP
fraction is difficult at the moment. However, it seems that the number 
of $20\%$ H-deficient post-AGB stars can not be produced from one
scenario alone. Thus, it will be a future challenge to
assign individual objects to a specific scenario.
\begin{table}[tbhp]
\caption{Classification of post-AGB evolution, : indicates uncertain
values.\label{tab:class}
}
  \begin{center}
    \begin{tabular}[h]{p{0.15\textwidth}p{0.39\textwidth}p{0.21\textwidth}}
\hline
Name & occurrence of H-deficiency & $X_{\rm H}$     \\
\hline
AFTP & during 1st AGB departure & $0.15$ to  solar (:)\\
VLTP & before return to 2nd AGB stage & $\leq 10^{-7}$ \\
LTP  & during 2nd AGB stage  & $\sim 0.02$ \\
\hline
    \end{tabular}  
  \end{center}
\end{table}

For example, Sakurai's object is
commonly believed to be a typical case of a Born-again star
(therefore no AFTP descendant)
witnessed during its rapid evolution. From our models it can
currently not be in the process of a LDUP. According to Asplund
(1999)  the object is in the second AGB phase,  already H-deficient (${\rm
X_H} < 1\%$) and the H-abundance is further decreasing. At the
same time the lithium abundance is increasing significantly. However,
during a LDUP the  Li-abundance increase is only expected during the
initial phase of DUP where the star is still H-normal (${\rm
X_H} \sim  65\%$). The following H depletion is correlated with a
lithium abundance decrease (Fig.\,\ref{fig:ABUND}) and does therefore
not resemble Sakurai's object.
The pattern of abundance change of this star  is closer to that found
during a  VLDUP where the convection
penetrates a zone which is H-free and Li-rich. However the observed
H-abundance - although already small - is still
larger by some orders of magnitude than that predicted by our current
VLTP model. Maybe this inconsistency can be resolved
by further investigation of VLTP models with progenitors
that depart from the AGB at a slightly different interpulse phase
compared to our current VLTP model.

\section{Appendix: The numerical method for coupled mixing and nuclear burning}

We have developed a numerical method for the fully coupled computation
of mixing and nuclear burning to be used in conjunction with a
stellar evolution code. It allows to follow the abundance change under 
conditions where the (convective) mixing time scale and the time scale of
nuclear burning are of the same order of magnitude.

Mixing in convectively unstable stellar regions (Langer \etal1985)
and in the adjacent overshoot regions (Herwig \etal1997)
can be mathematically described by a diffusion equation.
For each isotope $i$ ($i=1\,...\,i_{\rm max}$) a second order partial
differential equation of the form
\begin{equation}
\label{eq:misch}
\left(\frac{\mathrm{d} X_i}{\mathrm{d} t}\right)_{\rm mix}\,=\,
\frac{\partial}{\partial m}\left[\left(4\pi
r^2\rho\right)^2D\frac{\partial X_i}{\partial m}\right]
\end{equation}
is set up where $X_i$ is the abundance of isotope  $i$, $\rho$
density,
$r$ radius and $m$ mass coordinate.
The nuclear burning processes are defined by
a set of reaction rates, the nuclear network.
The time derivative of the isotope abundances at mass grid point  $j$
can be written as
\begin{equation}
\label{eq:brenn}
\left(\frac{\mathrm{d} \vec{X}_j}{\mathrm{d} t}\right)_{\rm burn}\,=\,
\hat{F}_j \cdot  \vec{X}_j  .
\end{equation}
The vector $\vec{X}_j$ contains the abundances of all considered
isotopes at grid point $j$. $\hat{F}_j$ is the rate matrix which
contains the functional dependence of the reaction rates
on the state variables. 
The overall abundance change over a given time step is the sum of the
contribution from mixing and from burning respectively (Eq.\,\ref{eq:misch} and
\ref{eq:brenn}).
%
%
It  is discretized fully implicit for the mixing term and
by a simple Euler step (likewise implicit) for the burning term.
The discritized equation can be written for each grid point $j$:
\begin{equation}
\label{eq:dikr1}
\vec{X}^{\mathrm{n+1}}_{j}-\vec{X}^{\mathrm{n}}_j
=h\hat{F}_j(\vec{X}^{\mathrm{n+1}}_j) 
+ h[\hat{P}_j \vec{X}^{\mathrm{n+1}}_{j+1} 
-\hat{Q}_j \vec{X}^{\mathrm{n+1}}_{j}
+\hat{R}_j \vec{X}^{\mathrm{n+1}}_{j-1} ]    ,
\end{equation}
where  $h$ is the time step, $n$ is the index numbering the time step
and $\hat{P}_j$, $\hat{Q}_j$ and $\hat{R}_j$ 
are diagonal matrices with the diagonal elements
\begin{eqnarray*}
  p_j= \frac{D^{*}_{\mathrm{j-1}}}{ml_{\rm j}} \frac{\Delta t}{mm_{\rm j}},
  r_j=\frac{D^{*}_{\rm j}}{mr_{\rm j}} 
  \frac{\Delta t}{mm_{\rm j}}   ,\\
  q_j= (\frac{D^{*}_{\rm j}}{mr_{\rm j}} + 
   \frac{D^{*}_{\rm j-1}}{ml_{\rm j}})\frac{\Delta
   t}{mm_{\rm j}}\\
  {\rm with\ } ml_{\rm j}=m_{\rm j}-m_{\rm j-1}
  ,{\rm \ }
  mr_{\rm j}=m_{\rm j+1}-m_{\rm j}
  ,\\
  mm_{\rm j}=\frac{1}{2}(m_{\rm j+1}-m_{\rm j-1})
  {\rm \ and\ }
  D^{*}_{\rm j}=\left(4\pi r^2\rho\right)^2 D_{\rm j}
\end{eqnarray*}
at the grid point  $j$.
Neighbouring grid points are coupled by the diagonal operators
$\hat{P}_j$ and $\hat{R}_j$. The discretization for the first and last grid point follows from the boundary condition of vanishing mass flux.

Equation \ref{eq:dikr1} has been solved by  Newton's method. 
For each grid point a vector function according to Eq.\,\ref{eq:dikr1} is given by:
\begin{equation}
\label{eq:G-in-Schale}
\vec{G}_j=\vec{X}^{\rm n+1}_j-\vec{X}^{\rm n}_j
-h\hat{F}_j(\vec{X}^{\rm n+1}_j) 
- h[\hat{P}_j \vec{X}^{\rm n+1}_{j+1} 
-\hat{Q}_j \vec{X}^{\rm n+1}_{j}
+\hat{R}_j \vec{X}^{\rm n+1}_{j-1} ]    .
\end{equation}
The independent variable of this non-linear function is the abundance
$\vec{X}^{\rm n+1}$ belonging to the time
$t^{\rm n+1}=t^{\rm n}+h$.
The grand vectors  $\vec{\mathbf{G}}$ and  $\mathbf{\vec{X}^{\rm
n+1}}$ contain $\vec{G}_j$ and $\vec{X}^{\rm n+1}_j$ for all grid
points $j$.
%
%

$\vec{\bf X}^{\rm n+1}$ contains a solution of Eq.\,\ref{eq:dikr1} for each 
grid point ${\bf j} $ if
\begin{equation}
  \label{gi-zero}
{\bf \vec{G}}({\bf \vec{X}}^{\rm n+1})={\bf \vec{0}}.
\end{equation}
Newton's method  for root finding replaces the nonlinear
Eq.\,\ref{gi-zero} by a linear equation for the corrections $\delta
{\bf \vec{X}}^{n+1,l}$ to the previous solution
${\bf \vec{X}^{\rm n+1,l}}$: 
\begin{equation}
\label{power-exp-G}
{\bf \vec{G}}({\bf \vec{X}}^{\bf \rm n+1,l}) 
+ \frac{\partial {\bf \vec{G}}({\bf \vec{X}}^{\rm n+1,l})}
       {\partial {\bf \vec{X}}^{\rm n+1,l}}
       \delta   {\bf \vec{X}}^{\rm n+1,l} ={\bf 0} .
\end{equation}
The initial guess may be
${\bf \vec{X}^{\rm n+1,l}}={\bf \vec{X}^{\rm n}}$.
The block tri-diagonal Jacoby matrix $ \frac{\partial
{\bf \vec{G}}({\bf \vec{X}}^{\rm n+1,l})} {\partial
{\bf \vec{X}}^{\rm n+1,l}}$ contains on its main diagonal the matrices
\begin{equation}
 \frac{\partial \vec{G}_{\rm j}}{\partial \vec{X}_{\rm j}^{\rm n+1}}= \hat{1}-h\hat{F}_j + h\hat{Q}_j ,
\end{equation}
as well as on the upper and lower diagonal the elements
\begin{equation}
 \frac{\partial \vec{G}_{\rm j}}
{\partial \vec{Y}_{\rm j+1}^{\rm n+1}}= 
-h\hat{P}_j {\rm \ , und\ } 
\frac{\partial \vec{G}_{\rm j}}{\partial
\vec{X}_{\rm j-1}^{\rm n+1}}= -h\hat{R}_j   .
\end{equation}
By recursive solution of Eq.\,\ref{power-exp-G} a better
approximation for ${\bf \vec{X}}^{\rm n+1}$
can be found iteratively:
\begin{equation}
  \label{iter-hen}
  {\bf \vec{X}}^{\rm n+1,l+1}={\bf \vec{X}}^{\rm n+1,l}+
  \delta   {\bf \vec{X}}^{\rm n+1,l}.
\end{equation}

For one model of the previously described VLTP sequence 
typically three to five iterations minimize the
largest relevant relative correction with $\delta X_{\rm
ij}^{\rm max,rel} < 10^{-3}$. 
The largest  absolute corrections which usually belong to the more
abundant isotopes then have the order of magnitude
$\delta X_{\rm ij}^{\rm max,abs} < 10^{-6} \dots 10^{-8}$.
In the computations we found the smallest proton abundance in the
burning region which can be 
represented to be $10^{-14}$ (mass fraction).
With a hydrogen abundance of  $10^{-14}$ the half life time of  \czw\
against destruction by proton capture is $\simeq 10^{5} {\rm yr}$.\\

{\sc Acknowledgments:}  I would like to thank T.\ Bl\"ocker, L.\ Koesterke, N.\ Langer and K.\ Werner
for many very helpful discussions.
 This work has been supported by the \emph{Deut\-sche
    For\-schungs\-ge\-mein\-schaft} through grant La\,587/16.

\end{document}